\documentclass[letter,twocolumn]{jpsj2} 

\usepackage{graphicx}
\usepackage{dcolumn}
\usepackage{bm}

\begin{document}

\title{Universal Scaling in the Dynamical Conductivity of 
Heavy Fermion Ce and Yb Compounds}    

\author{%
Hidekazu \textsc{Okamura}\thanks{E-mail: okamura@kobe-u.ac.jp}, 
Tatsuya \textsc{Watanabe}$^1$, 
Masaharu \textsc{Matsunami$^1$}, 
Tomohisa \textsc{Nishihara$^1$}, 
Naohito \textsc{Tsujii$^2$}, 
Takao \textsc{Ebihara$^3$}, 
Hitoshi \textsc{Sugawara$^4$}, 
Hideyuki \textsc{Sato$^5$}, 
Yoshichika \textsc{\={O}nuki$^6$}, 
Yosikazu \textsc{Isikawa$^7$}, 
Toshiro \textsc{Takabatake$^8$} and 
Takao \textsc{Nanba$^1$}
}
\inst{%
$^1$Graduate School of Science and Technology, 
Kobe University, Kobe 657-8501
\\
$^2$National Institute for Materials Science, 
Tsukuba 305-0047
\\
$^3$Department of Physics, Faculty of Science, 
Shizuoka University, Shizuoka 422-8529
\\
$^4$Faculty of Integrated Arts and Sciences, 
University of Tokushima, Tokushima 770-8502
\\
$^5$Graduate School of Science, Tokyo 
Metropolitan University, Tokyo 192-0397 
\\
$^6$Graduate School of Science, Osaka University, 
Toyonaka 560-0043
\\  
$^7$Department of Physics, Toyama University, 
Toyama 930-8555 
\\
$^8$Graduate School of Quantum Matter, Hiroshima 
University, Higashi-Hiroshima 739-8526
}

\recdate{\today}

\abst{
Dynamical conductivity spectra [$\sigma(\omega)$] have been 
measured for many heavy-fermion (HF) Ce and Yb compounds.       
A characteristic excitation peak has been observed in the 
infrared region of $\sigma(\omega)$ for all the compounds, 
and has been analyzed in terms of a simple model based on 
conduction ($c$)-$f$ electron hybridized band.     
A universal scaling is found between the observed 
peak energies and the estimated $c$-$f$ hybridization 
strengths of these HF compounds.  
This scaling demonstrates that the model of $c$-$f$ hybridized 
band can generally and quantitatively describe the low-energy 
charge excitations in a wide range of HF compounds.  
}

\kword{Heavy fermion, optical conductivity}

\maketitle

Physics of ``heavy fermion (HF)'' compounds has attracted a 
considerable amount of interest during the last few decades.  
In HF compounds, a strong Coulomb correlation of the $f$ 
electrons enhances the carrier effective mass ($m^\ast$).   
The mass enhancement is manifested in, e.g., the quadratic 
temperature ($T$) coefficient ($A$) of the resistivity and 
the linear $T$ coefficient ($\gamma$) of the electronic 
specific heat \cite{hewson}.    
It has been shown that a scaling between $A$ and $\gamma^2$, 
now widely known as the ``Kadowaki-Woods (KW) relation'', 
holds for a large number of HF compounds having different 
degrees of mass enhancement and orbital degeneracy 
\cite{KW,tsujii}.    
Theoretically, the KW relation including effects of electron 
correlation was successfully derived using the periodic 
Anderson model (PAM),\cite{yamada} and later generalized to 
include effects of orbital degeneracy.\cite{kontani,tsujii}   
In addition to the interesting transport, magnetic, and thermal 
properties of HF compounds, their charge excitation spectra, 
measured as the dynamical (optical) conductivity $\sigma(\omega)$, 
have been shown to be highly anomalous compared with those of 
conventional metals.\cite{sievers,degiorgi}    
Such anomalies should arise from their peculiar electronic 
dispersion near the Fermi level ($\epsilon_F$), 
which is the main interest of this Letter.

The simplest case of a HF compound is that of a Ce (Yb) 
compound with a 4$f^1$ (4$f^{13}$) configuration, where 
a 4$f$ electron (hole) interacts with the conduction 
electrons.\cite{hewson}     
Using the Fermi liquid theory and PAM, it has been shown 
that the dispersion in such a case may be approximated 
in terms of ``renormalized hybridized 
bands'', as sketched in Fig.~1(a) \cite{Rice,Anderson,Cox}.     
They result from a conduction ($c$)-$f$ electron hybridization 
renormalized by the electron correlation.     
%
\begin{figure}
\begin{center}
\includegraphics[width=0.38\textwidth]{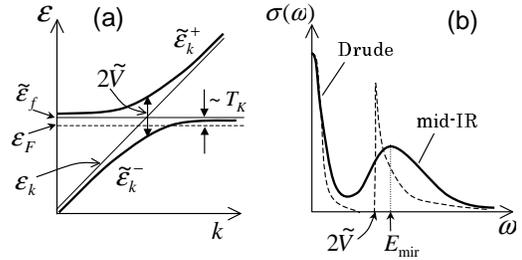}
\end{center}
\caption{
(a) Dispersion of the renormalized hybridized bands near 
$\epsilon_F$, expressed as 
$\tilde{\epsilon}_k^\pm=\frac{1}{2}[\tilde{\epsilon_f}+\epsilon_k 
\pm \sqrt{(\tilde{\epsilon_f}-\epsilon_k)^2 + 4\tilde{V}^2}]$ 
\cite{Rice,Anderson,Cox}.    
Here $\tilde{\epsilon_f}$ and $\tilde{V}$ are renormalized 
$f$ level and $c$-$f$ hybridization, respectively.  
$\epsilon_k$ is the bare $c$ band dispersion and $T_{\rm K}$ is 
the Kondo temperature.    
(b) Sketch of $\sigma(\omega)$ expected from 
$\tilde{\epsilon}_k^\pm$ \cite{Millis-Lee,Coleman}.    
The dashed and solid curves show those without and with 
disorder-induced broadening, respectively.       
}
\end{figure}
%
The $\sigma(\omega)$ spectrum resulting from this dispersion consists 
of a very narrow Drude peak centered at $\omega$=0 
\cite{Millis-Lee} and a peak at finite energy 
due to interband transitions [Fig.~1(b)] \cite{Coleman}.    
Experimentally, $\sigma(\omega)$ of several HF compounds 
exhibited a narrow Drude peak \cite{degiorgi} and a mid-infrared 
(mid-IR) absorption peak \cite{garner,basov,degiorgi-MIR,schle2}, 
which were consistent with the predictions in Fig.~1(b).     
Garner {\it et al} \cite{garner} first pointed out that, 
according to results of $c$-$f$ hybridized bands \cite{Cox}, 
the position 
(energy) of the mid-IR peak in $\sigma(\omega)$, $E_{\rm mir}$, 
should obey 
%
\begin{equation}
E_{\rm mir} \propto \sqrt{T_K W}, 
\end{equation}
%
where $W$ is the bare $c$ band width.     
Degiorgi {\it et al.} \cite{degiorgi-MIR} also discussed 
$E_{\rm mir}$ of HF 
compounds using Eq.~(1), while Dordevic {\it et al.} \cite{basov} 
reported a scaling of $E_{\rm mir}$ for a few HF compounds, 
using a relation essentially equivalent to Eq.~(1).     
Subsequently, Hancock {\it et al.} \cite{schle2} used Eq.~(1) 
to analyze $\sigma(\omega)$ spectra of YbIn$_{1-x}$Ag$_x$Cu$_4$.    
They showed that $E_{\rm mir}$ indeed scaled with 
$\sqrt{T_K}$ for 0 $\leq x \leq$ 1, establishing a 
scaling of $E_{\rm mir}$ in terms of Kondo physics 
for YbIn$_{1-x}$Ag$_x$Cu$_4$.     
However, since these previous works dealt with a limited range 
of HF compounds, it was unclear whether such a scaling of 
mid-IR peak based on $c$-$f$ hybridized band was universal 
for many HF compounds.

In this Letter, $\sigma(\omega)$ spectra of many Ce- and Yb-based 
HF compounds are systematically examined 
to test whether the mid-IR peaks follow a universal law among 
these compounds, and to clarify the above question.     
$\sigma(\omega)$ of CeNi, CeSn$_3$, YbInCu$_4$, YbAl$_2$, 
YbCu$_2$Si$_2$, YbNi$_2$Ge$_2$, and YbCuAl have been newly 
measured.     
$\sigma(\omega)$ spectra are also taken from our previous works 
on CeNiSn \cite{okamura-CeRhX}, CeRhSb \cite{okamura-CeRhX}, 
CeOs$_4$Sb$_{12}$ \cite{matunami-1}, 
CeRu$_4$Sb$_{12}$ \cite{matunami-2} and 
YbAl$_3$ \cite{okamura-YbAl3}.      
In addition, several other HF compounds for which $E_{\rm mir}$ 
values have been reported in the literature are considered.    
%
%
\begin{table}[b]
\begin{center}
\caption{
Specific heat coefficients ($\gamma$) of the Yb (Ce) compounds 
and those ($\gamma_0$) of the isostructural Lu (La) compounds 
in units of mJ/K$^2$mol.    The $\gamma$ values marked with 
asterisks are not from the cited references, but from our 
own analysis to separate a NFL contribution from the total 
$\gamma$ reported in the cited references.\cite{steglich,ISSP}    
}
\begin{tabular}{c|c|c||c|c|c} 
  & $\gamma$  & $\gamma_0$ &  &  $\gamma$ & $\gamma_0$   \\ \hline \hline
YbAl$_2$ &  17 \cite{tsujii-2}  &  5.6 \cite{LuAl2} & 
CeSn$_3$   &  53 \cite{onuki-CeSn3} & 2.5 \cite{LaSn3}  \\ 
\hline
YbAl$_3$  &  58 \cite{ebihara}  &  4 \cite{LuAl3}  &
CePd$_3$  &  38.6 \cite{CePd3}  &  3.5 \cite{YPd3}  \\ 
\hline
YbCu$_2$Si$_2$  &  135 \cite{tsujii-2} &  3 \cite{LuCu2Si2}  & 
CeRhSb    &  23-28 \cite{takaba} &  8.2 \cite{takaba}  \\ 
\hline
YbNi$_2$Ge$_2$  &  136 \cite{tsujii-2} & 12.7 \cite{LuNi2Ge2}  &
CeNi     &  57 \cite{onuki-CeNi}  &  5 \cite{onuki-CeNi} \\ 
\hline
YbCuAl   &  260 \cite{tsujii-2} & 5.9 \cite{LuCuAl}  &
CeNiSn   &  51 \cite{takaba}  &  11.4 \cite{takaba}  \\ 
\hline
YbInCu$_4$         &  50 \cite{cornelius}  &  8.7 \cite{lawrence}  & 
CeRu$_4$Sb$_{12}$  &  50-85$^\ast$ \cite{ISSP} &  37 \cite{ISSP}   \\ 
\hline
YbAgCu$_4$     &  210 \cite{cornelius} &  10 \cite{cornelius}  & 
CeOs$_4$Sb$_{12}$  &  92 \cite{maple2} & 36 \cite{maple2}  \\ 
\cline{1-3}
YbRh$_2$Si$_2$ & 100-   & 7.8\cite{LuRh2Si2} 
& & 180 \cite{metro} & \\ 
\cline{4-6} 
 & 140$^\ast$ \cite{steglich} &   & CeCu$_6$   &  
1600 \cite{wachter}  &  8 \cite{LaCu6}  \\
\hline
\end{tabular} 
\end{center}
\end{table}
%
Table~I lists the physical parameters of the studied HF compounds, 
used below to evaluate their $T_{\rm K}$ and $W$.     
Note that these compounds span a wider range of $\gamma$ 
values compared with previous works,\cite{degiorgi,basov,schle2} 
which is important for our purpose.       
It is shown that their $E_{\rm mir}$ values scale well 
in the form of Eq.~(1).

The newly measured samples were polycrystals for YbAl$_2$ 
and YbCuAl \cite{tsujii-2}, and single crystals for CeNi 
\cite{onuki-CeNi}, CeSn$_3$ \cite{onuki-CeSn3}, YbInCu$_4$ 
\cite{ebihara-YbInCu4}, YbNi$_2$Ge$_2$ \cite{tsujii-2} and 
YbCu$_2$Si$_2$ \cite{tsujii-3}.     The optical reflectivity 
spectra [$R(\omega)$] were measured on as-grown specular 
surfaces for YbCu$_2$Si$_2$ and YbNi$_2$Ge$_2$, on cleaved 
surfaces for CeNi and YbInCu$_4$,\cite{okamura-YbInCu4} 
and on polished surfaces for the others.    
$\sigma(\omega)$ spectra were obtained from 
$R(\omega)$ using the Kramers-Kronig analysis.   Other details 
were similar to those described elsewhere.\cite{okamura-YbAl3}

Figure~2 shows the obtained $\sigma(\omega)$ spectra at 9~K.   
\begin{figure}[b]
\begin{center}
\includegraphics[width=0.425\textwidth]{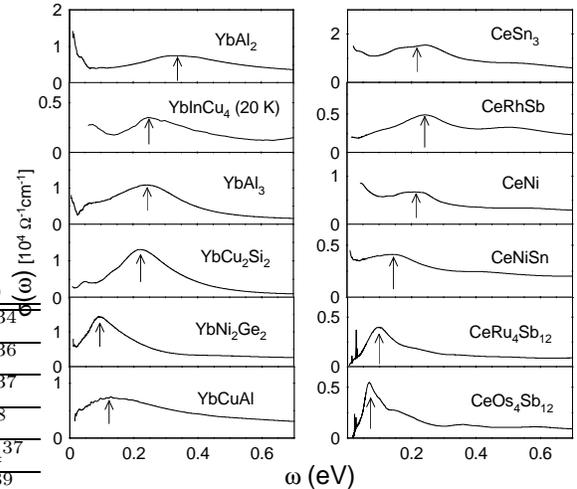}
\end{center}
\caption{Measured $\sigma(\omega)$ at 9~K, except for YbInCu$_4$ 
measured at 20~K for technical limitation.   The arrows indicate 
the $E_{\rm mir}$ values (peak positions) of the mid-IR peaks.  
}
\end{figure}
Since the dispersion sketched in Fig.~1 does not include thermal 
effects, here we discuss low-$T$ spectra only.  
All these compounds clearly exhibit a mid-IR peak in 
$\sigma(\omega)$, which strongly suggests that the 
appearance of a mid-IR peak is a universal property 
of the HF compounds.    
Their $E_{\rm mir}$'s (peak positions), indicated by 
the arrows in Fig.~2, vary markedly from compound to compound, 
which indicates a wide range of their $\tilde{V}$ and $T_K$.    
Our $\sigma(\omega)$ spectra of CeSn$_3$ and CeRu$_4$Sb$_{12}$ 
are similar to those in Refs.~\cite{bucher,basov}, although 
the overall magnitude of our $\sigma(\omega)$ is larger.     
Our $E_{\rm mir}$ of 0.25~eV for YbInCu$_4$, measured on a 
cleaved sample, agrees with the previously reported 
value\cite{schle2} for a polished sample, while the overall 
magnitude of our $\sigma(\omega)$ is much 
smaller.\cite{okamura-YbInCu4}

Eq.~(1) can be derived from the relation 
$2\tilde{V} \propto \sqrt{T_{\rm K} W}$ \cite{Cox}, which is a 
result of PAM, and by setting $E_{\rm mir} \propto 2\tilde{V}$.   
Although $E_{\rm mir}$ is larger than 2$\tilde{V}$ as sketched 
in Fig.~1(b), Hancock {\it et al.} \cite{schle2} have well 
demonstrated that $E_{\rm mir} \propto 2\tilde{V}$ 
holds in fact for YbIn$_{1-x}$Ag$_x$Cu$_4$ (0 $\leq x \leq$ 1).      
To analyze the measured $E_{\rm mir}$ values using Eq.~(1), 
it should be expressed in terms of experimentally observable 
quantities.   
$W$ of a Ce (Yb) compound may be regarded as inversely 
proportional to $\gamma$ of the non-magnetic, isostructural 
La (Lu) compound (denoted as $\gamma_0$).   
Concerning $T_{\rm K}$, since we need to deal with a large number 
of compounds, we shall make use of the simple relation\cite{hewson} 
%
\begin{equation}
T_{\rm K}=\pi^2 N_{\rm A} k_{\rm B} a/(3\gamma).   
\end{equation}
%
Here, $N_{\rm A}$ and $k_{\rm B}$ are the Avogadro number and 
Boltzman constant, respectively, and $a$ is a constant which 
depends only on the $f$ level degeneracy $N$: $a$=0.21, 0.54 
and 0.59 for $N$=2, 6 and 8, respectively \cite{hewson}.    
Then Eq.~(1) can be rewritten as 
%
\begin{equation}
E_{\rm mir} \propto \sqrt{a/(\gamma \gamma_0)}.  
\end{equation}
%
The constant of proportionality in Eq.~(3) is, in the absence 
of broadening ($E_{\rm mir}=2\tilde{V}$), given as
%
\begin{equation}
S_{\rm th}=\sqrt{8 \pi^3 / 9} \cdot N_{\rm A} k_{\rm B}^2, 
\end{equation} 
%
which is material independent.

In analyzing the experimental $E_{\rm mir}$ values using Eqs.~(3) 
and (4), we also include reported $E_{\rm mir}$'s in the literature 
for YbAgCu$_4$ \cite{schle2}, YbRh$_2$Si$_2$ \cite{kimura2}, 
CePd$_3$ \cite{bucher} and CeCu$_6$ \cite{wachter}.    
Except for CeRu$_4$Sb$_{12}$ \cite{ISSP} and 
YbRh$_2$Si$_2$,\cite{steglich} the HF compounds have Fermi 
liquid (FL) ground state at low $T$.    
Their $\gamma$ values in Table~I are taken from the cited 
references, which were determined through the usual procedure, 
i.e., by extrapolating the measured $C_{\rm e}/T$ to $T$=0 
($C_{\rm e}$ is the electronic specific heat).   
The $\gamma$ value thus obtained is the ``Kondo contribution'' 
proportional to $1/T_K$ as discussed above.    
CeRu$_4$Sb$_{12}$ \cite{ISSP} and YbRh$_2$Si$_2$ \cite{steglich} 
show steep rise in $C_e/T$ due to non-Fermi liquid (NFL) properties 
below $\sim$ 4~K and $\sim$ 10~K, respectively.     
Such a NFL effect is not included in our simple model of Eq.~(2).  
However, since the onset $T$ of this ``NFL contribution'' is clearly 
identified in their $T$ dependence of $C_e/T$ \cite{ISSP,steglich}, 
it can be subtracted from the total $C_e/T$ to estimate a 
Kondo contribution,\cite{ISSP,steglich} as previously done for, 
e.g., another NFL compound CeRhSn \cite{CeRhSn}.    
(Of course, there is no rigorous theoretical justification for this 
subtraction, since the NFL properties should also arise from the 
strong electron correlation.   However, it is a simple, 
phenomenological scheme to estimate the underlying Kondo 
coupling of an NFL compound.\cite{CeRhSn})  
Their $\gamma$'s in Table~I are the Kondo contributions thus 
obtained \cite{ISSP,steglich}.     
Estimated uncertainty resulting from the subtraction is indicated 
by the horizontal error bars in Fig.~3, which are rather small 
since not $\gamma$ but $\sqrt{\gamma}$ enters Fig.~3.        
CeOs$_4$Sb$_{12}$ shows a gradual increase of resistivity 
below 50~K, although other physical properties show metallic 
characteristics including the $C_{\rm e}/T$ data 
\cite{maple2,metro}.    
CeNiSn and CeRhSb are metals having low carrier densities ($n$) 
and a reduced density of states (dos) at $\epsilon_{\rm F}$ 
\cite{takaba}.    
The reported $E_{\rm mir}$'s of YbFe$_4$Sb$_{12}$ 
\cite{basov,kimura} and CeCoIn$_5$ \cite{singley} are not 
included here, since YbFe$_4$Sb$_{12}$ is not an $f$ 
electron-derived HF \cite{kimura}, and since CeCoIn$_5$ 
shows strong NFL properties up to high $T$ \cite{CeCoIn5-NFL}, 
making it difficult to estimate the Kondo contribution to 
$\gamma$.

Figure~3(a) shows plots of the measured and reported 
$E_{\rm mir}$ values as a function of 
$\sqrt{a/(\gamma \gamma_0)}$ using the $\gamma$ and 
$\gamma_0$ values in Table~I.     
%
\begin{figure}
\begin{center}
\includegraphics[width=0.40\textwidth]{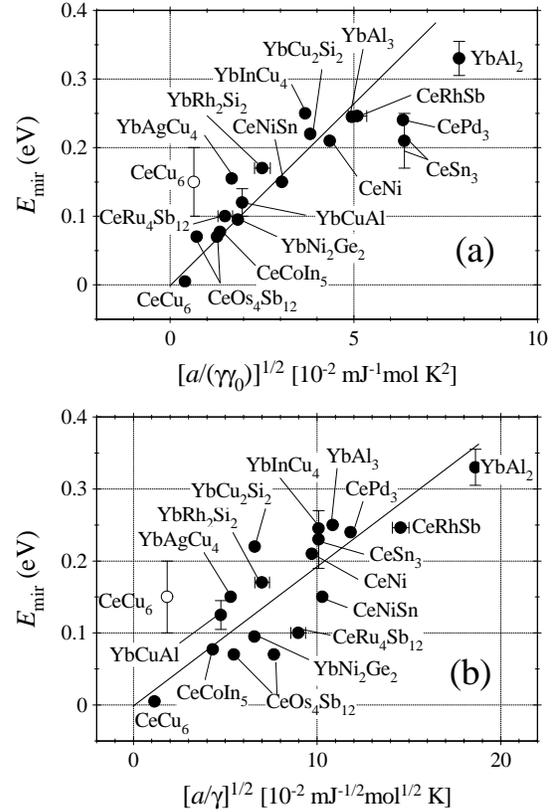}
\end{center}
\caption{
Mid-IR peak energies ($E_{\rm mir}$) plotted as a function of 
(a) $\sqrt{a / (\gamma \gamma_0)}$  and 
(b) $\sqrt{a / \gamma}$.   
The solid lines are guide to the eye.     The horizontal 
error bars indicate the uncertainties in $\gamma$ and the 
vertical ones those in determining $E_{\rm mir}$ due to 
broad lineshapes.     
}
\end{figure}  
Here, $N$=8 (6) has been used for the Yb (Ce) compounds, 
except for CeCu$_6$ (see below).      
Figure~3(a) shows that the $E_{\rm mir}$ values scale well, in 
agreement with the prediction of Eq.~(3).     
Some of the compounds, e.g., YbAl$_2$, CePd$_3$, CeSn$_3$ and 
CeCu$_6$, seem to show larger deviations than the others.    
However, considering the extreme simplicity of our model, 
the overall scaling is still good.    
The reported $\sigma(\omega)$ of CeCu$_6$ at low $T$ shows two 
marked IR peaks: a broad one centered at $\sim$ 0.15~eV and the 
other at 5~meV \cite{wachter}.    
Since the energy scale of 0.15~eV (1700~K) is much larger 
than $T_K$ of CeCu$_6$ ($\sim$ 1~K),\cite{wachter} the 5~meV 
peak rather than the 0.15~eV peak should be associated with 
the hybridization gap, as previously done.\cite{basov}    
Hence only the 5~meV peak is plotted in Fig.~3(a) using $N$=2 
since the Ce ground state in CeCu$_6$ is a $N$=2 doublet due 
to crystal field splitting.   (See the discussion below 
regarding the lowering of degeneracy in other compounds.)    
From Fig.~3(a), the experimental slope of the scaling is 
estimated as $S_{\rm ex}$ = (7 $\sim$ 9)~$\times$ 
10$^{-22}$ J$^2$/K$^2$mol, which is in excellent agreement 
with the theoretical slope given by Eq. (4), 
$S_{\rm th}$=6.0 $\times$ 10$^{-22}$ J$^2$/K$^2$mol.        
(Note that the actual 2$\tilde{V}$ is slightly smaller 
than $E_{\rm mir}$, as mentioned before.)   
For comparison, in Fig.~3(b) $E_{\rm mir}$ is plotted as a 
function of $\sqrt{a/\gamma}$ ($\propto \sqrt{T_{\rm K}}$), 
without including $\gamma_0$ ($\propto W^{-1}$).     
Although an overall tendency for scaling is still seen, 
it is much less satisfactory than that in Fig.~3(a).   This is 
in contrast to the previous result of good scaling without 
$W$ observed for YbIn$_{1-x}$Ag$_x$Cu$_4$ \cite{schle2}.   
Clearly, the explicit inclusion of $W$ ($\propto \gamma_0^{-1}$) 
in our analysis has been crucial for observing the good scaling 
over many HF compounds.    
Note that CeNiSn and CeRhSb show good scaling of $E_{\rm mir}$ 
despite their small $n$ \cite{takaba}, while they are 
completely off the KW relation \cite{tsujii}.    
These contrasting results are reasonable, since $S_{\rm th}$ 
does not depend on $n$, while the slope of KW relation does 
depend on $n$ \cite{tsujii}.   
Figure~3(a) convincingly demonstrates that the universal relation 
predicted by Eqs.~(2) and (3) is in fact followed by these HF 
compounds, and that the model of $c$-$f$ hybridized band has 
captured the essential physics involved in the low-energy 
dynamical conductivity of HF compounds.

The Ce or Yb ground state is a doublet also in CeNiSn 
\cite{takaba}, YbCu$_2$Si$_2$ \cite{tsujii} and YbRh$_2$Si$_2$ 
\cite{steglich}.     
Hence the $T_{\rm K}$ estimated from their $\gamma$ at low 
$T$ correspond to that of the $N$=2 ground state.    However, 
except for the 5~meV peak for CeCu$_6$ \cite{wachter}, their 
mid-IR peaks have a large energy scale of the order of 1000~K.   
Hence the mid-IR peaks should be associated with the 
``high-temperature $T_{\rm K}$'' ($T^h_K$) for the entire 
$J$=5/2 (7/2) multiplet with $N$=6 (8).    Since $T^h_K$ is 
larger than the ground state $T_{\rm K}$, our use of $\gamma$ 
at low $T$ should underestimate $T^h_K$ relevant to the mid-IR 
peak.   In spite of this, these compounds are scaled well 
in Fig.~3(a).    This is probably because our use of $N$=6 (8) 
has partly compensated for the underestimation of $T^h_K$: 
in the relation $T_{\rm K} \propto a/\gamma$, $a$=0.54 
(0.59) for $N$=6 (8) is greater than $a$=0.21 for $N$=2 
\cite{hewson}.    It is also remarkable that the two NFL 
compounds, YbRh$_2$Si$_2$ and CeRu$_4$Sb$_{12}$, scale 
reasonably well, once the Kondo contribution is separated 
from the total $C_e/T$.

Regarding the $c$-$f$ hybridized band model in Fig.~1, it should 
be recalled that only one linear combination of 4$f$ orbitals can 
hybridize with one conduction band per spin \cite{Rice,Anderson}. 
When the 4$f$ ground state is orbitally degenerate, as in many 
of the compounds studied here, the rest of the degenerate 4$f$ 
orbitals should remain unhybridized and dispersionless at 
$\tilde{\epsilon}_f$.\cite{kontani2}    
They should cause a sharp $f$-dos at $\tilde{\epsilon}_f$, 
which gives rise to, e.g., a large Van Vleck susceptibility 
in HF compounds \cite{kontani2}.    
With such an $f$-dos remaining at $\tilde{\epsilon}_f$, one 
might expect a more complicated dispersion than that in Fig.~1, 
and hence a more complicated $\sigma(\omega)$ spectrum.  
(A detailed discussion concerning effects of $f$ level 
degeneracy on the $c$-$f$ hybridized state has been 
given.\cite{saso})   
However, note that such an unhybridized $f$ level, which is 
completely uncoupled from the $c$ electron, cannot contribute 
to $\sigma(\omega)$ \cite{theory}.      
This is most likely the reason why the basic features 
of the mid-IR peak in $\sigma(\omega)$ can be understood 
without taking the 4$f$ orbital degeneracy into account.

The limitation of the present scaling is clear: it cannot 
deal with the U-based HF compounds having a $f^2$ or $f^3$ 
configuration.    Although there have been many reports of 
$\sigma(\omega)$ on U compounds \cite{degiorgi}, the 
hybridized band model of Fig.~1 based on $f^1$ configuration 
is not directly applicable to the U compounds.    
To quantitatively understand $\sigma(\omega)$ of U compounds, 
further progress is required in the theoretical understanding 
of their electronic dispersion.

In summary, it has been shown that the peak energy of the 
characteristic IR excitation in $\sigma(\omega)$ universally 
scales with the $c$-$f$ hybridization strength, 
estimated by $\sqrt{a/(\gamma \gamma_0)}$, for many Ce- and 
Yb-based HF compounds.    
This optical scaling demonstrates that the simple model of 
$c$-$f$ hybridization band can quantitatively account for 
the low-energy charge dynamics in these HF compounds.    
Of course, there are exceptions to the present scaling.    
For example, CeAl$_3$ does not exhibit an IR 
peak,\cite{degiorgi} although its $\gamma$ value is 
comparable to that of CeCu$6$.       
Such an exception is not surprising; even the KW relation, 
now widely regarded as being universal, breaks down for 
certain HF compounds.\cite{tsujii}   
The important point is that most of the Ce- and Yb-based 
HF compounds whose $\sigma(\omega)$ data are currently 
available have been considered here, and that they have 
shown the good scaling.     In this sense, we believe 
that the present optical scaling for the Ce and Yb 
compounds should be a universal property.

H. O. would like to thank Professors T. Mutou, H. Kontani, 
H. Harima, and T. Saso for useful discussions.     
Experiments were partly performed at UVSOR BL7B as a 
Joint Studies Program of IMS, and at SPring-8 BL43IR 
under the approval by JASRI (2004A0778-NSa-np).    
This work has been partly supported by COE Research 
(No. 13CE2002), and Priority Areas ``Skutterudites'' 
(Nos. 15072204, 15072206, 15072207) and 
``High Field Spin Science in 100~T'' (No.~451) from 
the Ministry of Culture, Education, Sports, Science 
and Technology.  


\end{document}